\documentclass[aps,pra,showpacs,superscriptaddress]{revtex4-1}
\usepackage{graphics}
\usepackage{graphicx}
\usepackage{dcolumn}
\usepackage{bm}
\usepackage{amssymb,amsmath}

\begin{document}

\title{Modelling spontaneous four-wave mixing in periodically-tapered waveguides}

\author{Mohammed F. Saleh}

\affiliation{Scottish Universities Physics Alliance (SUPA), Institute of Photonics and Quantum Sciences, Heriot-Watt University, EH14 4AS Edinburgh, UK}
\affiliation{Department of Mathematics and Engineering Physics, Alexandria University, Alexandria, Egypt}

\email{m.saleh@hw.ac.uk} 

\begin{abstract}
Periodically-tapered-waveguides technique is an emerging potential route to establish quasi-phase-matching schemes for efficient on-demand parametric interactions in  third-order nonlinear materials. In this paper, I investigate this method in enhancing spontaneous photon-pairs emission in fibres and planar waveguides with sinusoidally-varying cross sections. I have developed a general robust  quantum model to study this process under continuous or  pulsed-pump excitations. The model shows a great enhancement in photon-pairs generation  in waveguides with a small number of tapering periods that are feasible via the current fabrication technologies. I envisage that this work will open a new area of research to investigate how the tapering patterns can be fully optimised to  tailor the spectral properties of the output  photons  in  third-order nonlinear guided structures.
\end{abstract}


\maketitle

\section{Introduction}
Spontaneous photon-pair  emission in nonlinear media has become the  standard  approach for preparing quantum states of light in labs due to its efficiency, ease of use, and operation at practical conditions \cite{Spring17}.  In second-order nonlinear media,  a mother photon can be spontaneously downconverted into two daughter photons in three-wave mixing  process. Whereas, two pump photons will combine  to generate two other  photons  in a spontaneous four-wave mixing (SFWM) parametric process in third-order nonlinear media. 

Efficient parametric wave-mixing  processes are inherently constrained by energy and momentum conservation (the latter is known as the phase-matching condition). Quasi-phase-matching schemes  have been successfully employed in second-order nonlinear media to enable  on-demand parametric processes, using  periodically-poled ferroelectric crystals \cite{Armstrong62,Fejer92}.  This technique has had a remarkable impact in the field of nonlinear and quantum optics; it eliminates the stringent constrains posed on the directions and polarisations of the interacting waves, and allows for  on-chip collinear co-polarised strong nonlinear interactions to take place.  Using this technique, highly-efficient  and  pure single-photon sources have been widely demonstrated in bulk and integrated configurations \cite{Eckstein11,Spring13,Graffitti18}.

Silica fibres and silicon-photonic waveguides are the practical and convenient platforms for generating single photons for quantum communications  and integrated-quantum  applications, respectively   \cite{McMillan09,Silverstone13}. Enabling SFWM  parametric processes in these third-order nonlinear  platforms can be achieved via exploiting the nearly-phase-matching regime, or balancing waveguide and material dispersions \cite{Agrawal07}.  Similar to second-order nonlinear media,  these  conventional methods poses restrictions, for instance,  the  frequency-separation between the interacting photons and the operating dispersion regime.  Other alternative techniques such as photonic crystal fibres \cite{Rarity05,Garay-Palmett07},  birefringent silica waveguides \cite{Spring13,Spring17}, microresonantors \cite{Vernon17}, symmetric and asymmetric directional couplers \cite{Dong04,Jones18} have shown the ability to circumvent these limitations to obtain highly-efficient  and  pure  single photons in third-order nonlinear media.

In analogue to periodically-poled structures,  periodically-tapered waveguides (PTWs)  have been also introduced to correct phase mismatching of third-order nonlinear interactions via  sinusoidally modulating the width of planar waveguides  or the core of microstructured fibres (coined as dispersion-oscillating fibres) \cite{Driscoll12,Mussot18}. The tapering period was initially restricted to the millimetre range in planar structures. Very recently, PTWs with tapering periods in the micrometer range  have been successfully realised in planar waveguides  in two different configurations with either modulating  the core or the cladding using the Photonic Damascene process \cite{Hickstein18}.  These structures have been mainly exploited in tuning dispersive-wave emission, controlling supercontinuum generation, and manipulating modulation instability.  On the theoretical side, I have presented a new analysis based on the Fourier-series to explain how PTWs can be employed as a complete analogue to periodically-poled  structures \cite{Saleh18}. In PTWs both the linear and nonlinear coefficients are longitudinally varying resulting in multiple Fourier components of the  phase-mismatching and nonlinear terms.  Opposite Fourier components of the two terms can simultaneously cancel each other, allowing the conversion efficiency of a parametric process to grow along the structure, if right combinations of the tapering period and  modulation amplitude of the waveguide are used.

In this paper, I have applied the PTW-technique to generate  photon pairs via SFWM processes at on-demand frequencies and with high spectral purity. The paper is organised as follows. In Sec. \ref{Sec2}, a robust quantum model have been developed to study this process in tapered waveguides for continuous and pulsed pump sources. This model has then  been applied to study SFWM processes in dispersion-oscillating fibres and width-modulated planar silicon-nitride  waveguides in Sec. \ref{Sec3}. Finally, my conclusions are summarised in Sec. \ref{Sec4}.

\section{Modelling SFWMs processes in PTWs}\label{Sec2}
Consider a SFWM parametric process in a single-mode third-order nonlinear waveguide.  I have developed a model follows an approach that determines the energy flux across a  transverse plane over a quantisation time $T$  in the Heisenberg picture \cite{Huttner90}. Initially, it is assumed that the pump is a  monochromatic wave and the waveguide is uniform. Later, the model will be generalised to take into account propagation in tapered waveguides, as well as having pulsed pump sources. Raman nonlinearity is neglected, for simplicity,  which is also the case in silicon-nitride waveguides \cite{Guo18} and hollow-core fibres filled by noble gases \cite{Saleh11a}. 

\paragraph{Continuous pump wave. } Assuming that the pump ($p$) is a strong undepleted monochromatic classical wave, its electric field can be described as
\begin{equation}
\mathcal{E}_p\left(x,y,z,t\right)= \mathbf{Re}\left\lbrace E_p\left(x,y,z,t\right)\right\rbrace ,
\end{equation}
where  $x,y$ are the transverse coordinates, $z$ is the longitudinal direction,  $t$ is the time, $\mathbf{Re}$ is the real part, $ E_p=A_{p} F_p\left(x,y\right) e^{-j\left(\omega_p t-k_pz\right)}$, $A_p$, $F_p$, $\omega_p$,  $k_p $  are the amplitude, transverse profile, angular frequency, and propagation constant of the pump wave,  respectively, $k_p= n_p \omega_p/c$, $n_p$  is the refractive index at the pump frequency, and $c$ is the speed of light in vacuum. Using Maxwell equations and applying slowly-varying approximation \cite{Agrawal07},  self-phase modulation effect that influences the pump propagation can be included by replacing  $k_p$ with  
\begin{equation}
\kappa_p=k_p\left[1+\frac{3\chi^{(3)}A_p^2} {8 \,n_p^2\, S_p}\iint\left|F_p\right|^4  dx dy\right],
\end{equation}
where $\chi^{(3)}$ is the nonlinear coefficient, and $S_p=\iint\left|F_p\right|^2  dxdy $ is the pump beam area.

\paragraph{Signal and idler photons.}
The electric field operators of the converted single photons (either signal or idler) can be written as 
\begin{equation}
\hat{ \mathcal{E}}\left(x,y,z,t\right)=\hat{ \mathcal{E}}^+\left(x,y,z,t\right)+\hat{ \mathcal{E}}^-\left(x,y,z,t\right), 
\end{equation}
where $\hat{ \mathcal{E}}^+$ and $ \hat{ \mathcal{E}}^-$ are the positive and negative frequency-components of the field,  $\hat{ \mathcal{E}}^-=\left(\hat{ \mathcal{E}}^+ \right)^\dagger$, and $\hat{ \mathcal{E}}^+$  can be expressed in terms of a superposition of frequency-dependent mode operators \cite{Huttner90},
\begin{equation}
\hat{ \mathcal{E}}^+\left(x,y,z,t\right)=\displaystyle\sum_s \sqrt{\frac{\hslash\omega_s}{2\epsilon_o cTn_s S_s}}F_s\left(x,y\right)\hat{a}\left(z,\omega_s\right)e^{-j\omega_s t},
\end{equation}
$F$,  $S$, $\omega$, and  $n$ have the same definitions as for the pump wave, $\hslash$ is the reduced Planck's constant, $\epsilon_o$ is the dielectric permittivity, and $\hat{a}$ is annihilation operator.  The discrete spectral modes are separated by a frequency spacing $2\pi/T$, and $T$ tends to $\infty$ in the continuous case. The photon-number operator of mode $s$ during a time window $T$  is $\hat{N}\left(z,\omega_s\right)=\hat{a}^\dagger\left(z,\omega_s\right)\hat{a}\left(z,\omega_s\right)$, with $\hat{a}^\dagger$  the creation operator.  The space evolution of the operators are governed by 
\begin{equation}
-j\hslash\frac{\partial  \hat{\mathcal{E}}}{\partial z}=\left[\hat{\mathcal{E}},\hat{G}\right],
\label{eq1}
\end{equation}
where $\hat{G}\left(z\right)=\iint\int_0^T\hat{g}\left(x,y,z,t\right)dt \,dxdy  $ is the momentum operator,  $\hat{g}=\hat{\mathcal{D}}^{+}\hat{\mathcal{E}}^{-}+H.c.$ is the momentum flux,  $\hat{\mathcal{D}}=\epsilon_o n^2 \hat{\mathcal{E}}+ \hat{\mathcal{P}}_\mathrm{NL}$ is the electric displacement field,  $\hat{\mathcal{P}}_\mathrm{NL}$ is the nonlinear polarisation, and $H.c.$ is the Hermitian conjugate.  In the linear regime (L) and  substituting the electric field and momentum operators in Eq. (\ref{eq1}), it can be easily shown that
\begin{equation}
\frac{\partial\hat{a}_s^{\left(\mathrm{L}\right)}}{\partial z}=j k_s \hat{a}_s\left(z\right),
\end{equation}
where the commutation relation $\left[\hat{a}\left(z,\omega_s\right),\hat{a}^\dagger\left(z,\omega_{s^{'}}\right)\right]=\delta_{ss{'}}$  has been applied, with $\delta$ a Kroenecker delta, $k_s = n_s\omega_s/c$, and $\hat{a}_s\left(z\right)=\hat{a}\left(z,\omega_s\right)$ for brevity.

 \paragraph{Spontaneous four-wave mixing process.}
To study a SFWM interaction in this picture,  the positive frequency-part of the nonlinear polarisation associated with this process should be first determined, $ \mathcal{\hat{P}_\mathrm{FWM}^+}=3\epsilon_o \chi^{(3)} E_pE_p\hat{ \mathcal{E}}^-$. Then, the momentum flux operator  is $\hat{g}_\mathrm{FWM}=\mathcal{\hat{P}_\mathrm{FWM}^+}\hat{\mathcal{E}}^-+H.c$. Applying the definition of the delta function $\delta\left(\Delta\omega\right) = \frac{1}{T} \int_0^T \exp\left(-j\Delta\omega t\right) dt$, the momentum operator in this case is determined as
\begin{equation}
\hat{G}_\mathrm{FWM}=\frac{3\hslash\chi^{(3)}A_{p}^2e^{j2\kappa_pz}}{8c}\sum_s \sqrt{\frac{\omega_s\omega_i}{n_s n_i S_s S_i}}\iint F_p^2 F_s^*F_i^*dx dy\; \hat{a}_s^{\dagger}\left(z\right)\hat{a}_i^{\dagger}\left(z\right)+H.c.\, ,
\end{equation}
with $\omega_i=2\omega_p-\omega_s$. The subscript $i$ refers to another mode that is coupled to the mode $s$ within the spectrum. Substituting $\hat{G}_\mathrm{FWM}$ in Eq. (\ref{eq1}), the space evolution of the mode operator $s$ due to this process is
\begin{equation}
\frac{\partial\hat{a}_s^{\left(\mathrm{FWM}\right)}}{\partial z}=j \frac{3\chi^{(3)}A_{p}^2e^{j2\kappa_pz}}{4c} \sqrt{\frac{\omega_s\omega_i}{n_s n_i S_s S_i}}\iint F_p^2 F_s^*F_i^*dx dy\;\; \hat{a}_i^{\dagger}\left(z\right).
\end{equation}

\paragraph{Cross-phase modulation (XPM) effect. } 
Following the aforementioned  procedure, the nonlinear part of the momentum flux operator due to the XPM effect is $\hat{g}_\mathrm{XPM}=6\epsilon_o \chi^{(3)}  E_p E_p^*\hat{\mathcal{E}}^+\hat{\mathcal{E}}^-+H.c$. Hence, the momentum operator of this effect is
\begin{equation}
\hat{G}_\mathrm{XPM}=\frac{3\hslash\chi^{(3)}A_p^2}{4c}\sum_s \frac{\omega_s}{n_s S_s }\iint \left|F_p\right|^2 \left|F_s\right|^2 dx dy\; \hat{a}_s\left(z\right)\hat{a}_s^{\dagger}\left(z\right)+H.c.\, ,
\end{equation}
 and the annihilation operator is   spatially evolved as 
 \begin{equation}
\frac{\partial\hat{a}_s^{\left(\mathrm{XPM}\right)}}{\partial z}=j\frac{3\chi^{(3)}A_p^2}{2c}\frac{\omega_s}{n_s S_s }\iint \left|F_p\right|^2 \left|F_s\right|^2 dx dy\; \; \hat{a}_s\left(z\right).
\end{equation}

 \paragraph{Total effects.}
Therefore, the  evolution of the mode operator $s$ due to linear and nonlinear contributions can be written as
 \begin{equation}
\frac{\partial\hat{a}_s}{\partial z}= \frac{\partial\hat{a}_s^{\left(\mathrm{L}\right)}}{\partial z}+\frac{\partial\hat{a}_s^{\left(\mathrm{XPM}\right)}}{\partial z}+\frac{\partial\hat{a}_s^{\left(\mathrm{FWM}\right)}}{\partial z}=j \kappa_s \hat{a}_s\left(z\right)+j\gamma_{s, i}\, e^{j2\kappa_pz}\hat{a}_i^{\dagger}\left(z\right),
\label{eq2}
\end{equation}
with 
 \begin{equation}
  \kappa_s = k_s\left( 1+ \frac{3\chi^{(3)}A_p^2}{2 n_s^2 S_s}\iint \left|F_p\right|^2 \left|F_s\right|^2 dx dy\right),
\end{equation}
and 
  \begin{equation}
 \gamma_{s, i}= \frac{3\chi^{(3)}A_{p}^2}{4c} \sqrt{\frac{\omega_s\omega_i}{n_s n_i S_s S_i}}\iint F_p^2 F_s^*F_i^*dx dy.
\end{equation}
A similar equation can be written for the mode operator $i$,
 \begin{equation}
\frac{\partial\hat{a}_i^{\dagger}}{\partial z}=-j \kappa^{*}_i\hat{a}_i^{\dagger} \left(z\right)-j\gamma^*_{s,i}\, e^{-j2\kappa_pz}\hat{a}_s\left(z\right).
\label{eq3}
\end{equation}
The two coupled Eqs. (\ref{eq2},\ref{eq3}) can be solved exactly in the case of uniform waveguides, similar to the spontaneous parametric downconversion process \cite{Huttner90}. 

 \paragraph{Transfer matrix method.}
Introducing the phase transformation $\hat{a}_q\left(z\right)=\hat{b}_q\left(z\right)\ e^{j\kappa_qz}$, $q=s,i$, Eqs. (\ref{eq2},\ref{eq3}) can be written in the common compact form, 
\begin{equation}
\frac{\partial\hat{b}_s}{\partial z}= j\gamma_{s, i}\, e^{j\Delta \kappa z}\hat{b}_i^{\dagger}\left(z\right),
\qquad 
\frac{\partial\hat{b}_i^{\dagger}}{\partial z}=-j\gamma^*_{s,i}\, e^{-j\Delta \kappa z}\hat{b}_s\left(z\right),
\end{equation}
with $\Delta \kappa =  2\kappa_p- \kappa_s-\kappa_i$. 
If the waveguide is  tapered, the coefficients in these equations become spatial dependent. Also,  the term $e^{j\Delta \kappa z}$ should be replaced by $e^{j\Delta \phi}$, with $\Delta \phi =\int_0^{z} \Delta \kappa\left(z'\right) z' \,dz'$   the proper accumulated  phase-mismatching \cite{Love91,Fejer92}. To solve the  coupled equations in this case, the waveguide can be truncated  into discrete infinitesimal elements with constant  cross sections, then the equations can be solved within each element. The outcome of an element $m$ can be written in a transfer-matrix form,
 \begin{equation}
 \left[ \begin{array}{c}
\hat{b}_s\\
\hat{b}_i^{\dagger}
\end{array}\right]_{z=z_m+\Delta z}= \mathcal{T}_m\left[ \begin{array}{c}
\hat{b}_s\\
\hat{b}_i^{\dagger}
\end{array}\right]_{z=z_m},\qquad
\mathcal{T}_m= 
\left[ \begin{array}{cc}
1&f_{s, i}\left(\omega_p\right)\\
f^*_{s, i}\left(\omega_p\right)&1
\end{array}\right]_{z=z_m},
\label{eq4}
\end{equation}
with $f_{s, i}\left(\omega_p\right)\ =j\gamma_{s, i} \Delta z\,e^{j\Delta \phi}$ and  $\Delta z$ the element thickness. The explicit dependence of the parameter $f_{s, i}$ on $\omega_p$ has been  shown here, to remind the reader that all these matrix elements depend on the pump frequency. The transfer matrix that describes the whole structure and relates the output operators to the input ones is given by the multiplication of the transfer matrices of all elements in a descending order, $\mathbf{T}_{s,i}=\mathcal{T}_M\mathcal{T}_{M-1}...\mathcal{T}_2\mathcal{T}_1$, with $M$  the elements number. It is worth to note that each element has a different transfer matrix even in periodic structures, because of the parameter $\Delta\phi$ that counts the accumulated phase from the beginning of the structure.

\paragraph{Expected number of photons.} The expected number of photons of a certain mode $s$ at the end of the  waveguide  is $\left<\psi\right|\hat{N}\left(L,\omega_s\right)\left|\psi\right>=\left<\psi\right|\hat{b}_s^{\dagger}\left(L\right)\hat{b}_s\left(L\right)\left|\psi\right>$, with  $\left|\psi\right>=\left|0\right>_s \left|0\right>_i$ the input quantum state and   $L$ the waveguide length. Using the transfer matrix of the whole structure, $\hat{b}_s\left(L\right)=\mathbf{T}_{s,i}\left(1,1\right)\hat{b}_s\left(0\right)+\mathbf{T}_{s,i}\left(1,2\right)\hat{b}_i^{\dagger}\left(0\right)$, and $\left<\psi\right|\hat{N}\left(L,\omega_s\right)\left|\psi\right>=\left|\mathbf{T}_{s,i}\left(1,2\right)\right|^2$. It is important to mention that this procedure is only for a single mode $s$ and it should be repeated for all other modes that compose the signal  spectrum.

\paragraph{Pulsed pump source.} The model that I have developed so far is based on having a monochromatic pump source. If the pump is assumed to be a Gaussian pulse with a characteristic temporal width $\tau$ and a central frequency $\omega_{p_{0}}$,  then $ E_p=A\left(t\right) F_p\left(x,y\right) e^{jk_pz}$, where $ A\left(t\right) = A_0 e^{-t^2/2\tau^2} e^{-j\omega_{p_{0}} t}$ is the pulse envelope and the full-width-half-maximum of this pulse equals $2\tau\sqrt{\ln(2)}$. Using the Fourier Transform, the  pump electric field can be decomposed into a superposition of multiple monochromatic waves with frequencies $\omega_p$,
 \begin{equation}
\mathcal{E}_p\left(x,y,z,t\right)= \mathbf{Re}\left\lbrace \sum_{\omega_{p}} A_p\left(\omega_p\right)\,e^{-j\left(\omega_p t-k_pz\right)}\right\rbrace ,
\end{equation}
where $A_p\left(\omega_p\right)=\frac{A_0 \tau \Delta \omega}{\sqrt{2\pi}}\,e^{-\frac{\tau^2\left(\omega_p-\omega_{p_{0}}\right)^2}{2}}$ is the amplitude of each component, and $ \Delta \omega$ is the sample frequency of the pump pulse.

To study how the evolution of two certain coupled modes $s$ and $i$ is influenced by having a pulsed pump source, each possible combination of two monochromatic pump waves $\omega_{p_{1}}$ and  $\omega_{p_{2}}$ in the  spectrum should be counted towards the evolution of the mode operators. Because of the energy conservation or the delta function used in determining the momentum operator $\hat{G}_\mathrm{FWM}$, the double summation over  $\omega_{p_{1}}$ and  $\omega_{p_{2}}$ is reduced to a single summation over  $\omega_{p_{1}}$ with $\omega_{p_{2}}=\omega_s+\omega_i-\omega_{p_{1}}$. The  above procedure to determine the expected number of photons can still be followed, however, the transfer matrix of an element $m$ becomes
 \begin{equation}
\mathcal{T}_m= 
\left[ \begin{array}{cc}
1&\displaystyle\sum_{\omega_{p_{1}}}f_{s, i}\left(\omega_{p_{1}},\omega_{p_{2}}\right)\\
 \displaystyle\sum_{\omega_{p_{1}}}f_{s, i}^*\left(\omega_{p_{1}},\omega_{p_{2}}\right)&1
\end{array}\right]_{z=z_u}.
\end{equation}
 For non degenerate combinations of $\omega_{p_{1}}$ and $\omega_{p_{2}}$,
 \begin{equation}
 \gamma_{s, i}= \frac{3\chi^{(3)}A_{p_{1}}A_{p_{2}}}{2c} \sqrt{\frac{\omega_s\omega_i}{n_s n_i S_s S_i}}\iint F_{p_{1}}F_{p_{2}} F_s^*F_i^*dx dy, 
\end{equation}
and $\Delta \kappa =  \kappa_{p_{1}}+\kappa_{p_{2}}- \kappa_s-\kappa_i$.  Also, each monochromatic pump wave in this case will induce a XPM effect that will influence the other pump wave  as well as the converted single photons. Therefore,
\begin{equation}
\kappa_{p_{u}}=k_{p_{u}}\left[1+\frac{3\chi^{(3)}} {8 \,n_{p_{u}}^2 S_{p_{u}}} \iint \left(A_{p_{u}}^2\left| F_{p_{u}}\right|^2+2A_{p_{v}}^2\left| F_{p_{v}}\right|^2\right)\left| F_{p_{u}}\right|^2   dx dy\right],\;\;
\left\lbrace
\begin{array}{l}
 u,v=1,2 \\
u\neq v 
\end{array},
\right. 
\end{equation}
and
\begin{equation}
 \kappa_q = k_q\left[ 1+ \frac{3\chi^{(3)}}{2 n_q^2 S_q}\iint \left(A_{p_{1}}^2 \left| F_{p_{1}}\right|^2  +A_{p_{2}}^2 \left| F_{p_{2}}\right|^2 \right)\left|F_q\right|^2  dx dy\right],\qquad q=s,i.
\end{equation}
Finally, it is more common in third-order nonlinear media to use the nonlinear refractive $n_2$ in units m$^2$/W rather than $\chi^{(3)}$ in units m$^2$/V$^2$. Using the definitions of $n_2$ and $\chi^{(3)}$ \cite{Agrawal07}, one can show that $3\chi^{(3)}A_{p}^2=8\,n_2I_{p}$, where $I_{p}=\frac{\mathbf{E}_i \tau \Delta \omega^2}{2\pi\sqrt{\pi}\,S_{p}} \, e^{-\tau^2\left(\omega_{p}-\omega_{p_{0}}\right)^2}$ is the intensity of a Fourier pump component $p$, and $\mathbf{E}_i$ is the input pulse  energy. Similarly, the quantity $3\chi^{(3)}A_{p_{1}} A_{p_{2}}$ can be written as $8\,n_2\sqrt{I_{p_{1}}I_{p_{2}}}$.

\section{Results}\label{Sec3}
In this section, the  developed model will be exploited in investigating photon-pair generation via SFWM parametric processes in periodically-tapered waveguides (PTWs), in particular, dispersion-oscillating fibres and width-modulated silicon-nitride waveguides. The platforms are designed to operate in the normal-dispersion regime, where satisfying the phase-matching condition is usually  hard to  achieve in  uniform single-mode waveguides. Operating in this regime will suppress unwanted nonlinear phenomena such as supercontinuum  generation, soliton formation,  or dispersive-wave emission that could ruin the aimed SFWM process under strong pumping \cite{Kowligy18}.  To insure adiabatic propagation, only small modulation amplitude and relatively-large tapering periods have been explored  \cite{Love91}. The tapering period in the simulated structures has been discretised into 200 steps to increase the accuracy of the computational results.

\subsection{Dispersion-oscillating fibres} 
\begin{figure}
\centerline{\includegraphics[width=14.5cm]{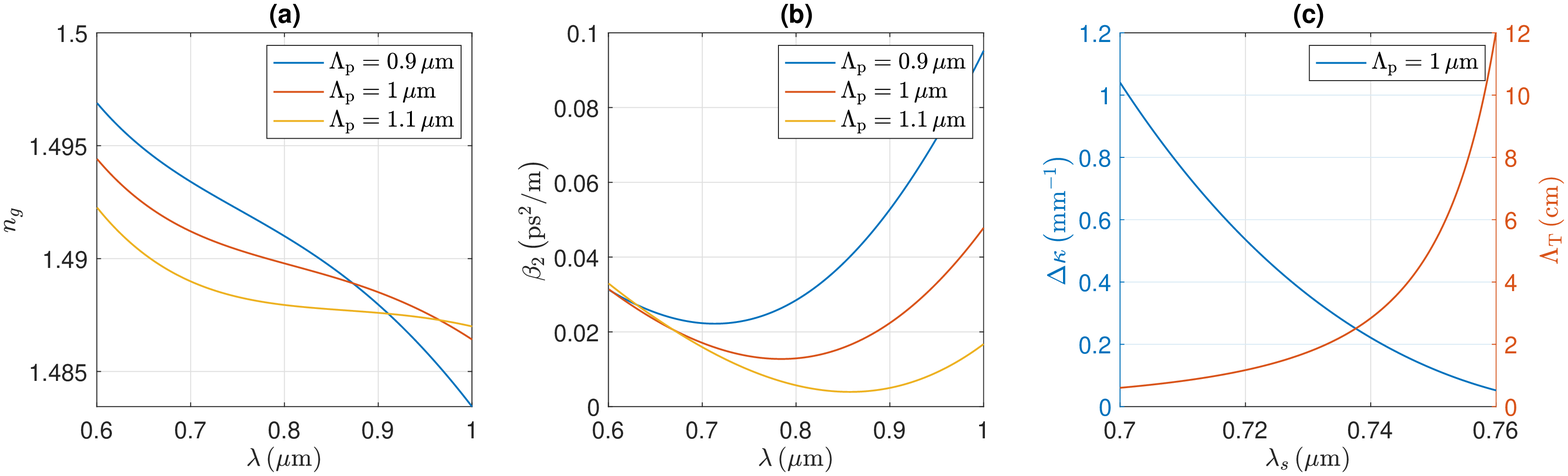}}
\caption{ (a,b) Wavelength-dependence of the group-index and second-order dispersion of a sinusoidally-tapered microstructured fibres with a pitch $\sigma$ varies between 0.9 and 1.1 $\mu$m, and a hole-diameter $d=0.5 \sigma$. (c) Dependence of the mismatching in the propagation constants $\Delta\kappa$ of a SFWM process and the  tapering period $\Lambda_\mathrm{T}$ on the signal wavelength $\lambda_s$ with a pump at 780 nm and an input power 1 W. The assumed fibre nonlinear refractive index is $n_2=2.25\times 10^{-20}$ m$^{2}$/W. These parameters will be used in subsequent simulations, unless stated otherwise.
\label{Fig1}}
\end{figure}

These are silica solid-core sinusoidally-tapered microstructured fibres made of a stack of hollow capillary tubes with a cross-sectional  pitch $\sigma$, and a hole-diameter $d$ \cite{Mussot18}. The  pitch varies as $\sigma(z)=\sigma_{\mathrm{av}}\left[1-\Delta\sigma\,\cos\left( 2\pi z/\Lambda_\mathrm{T}\right)\right]$, where $\sigma_{\mathrm{av}}$ is the average pitch, $\Delta\sigma$ is the modulation amplitude, and $\Lambda_\mathrm{T}$ is the tapering period.  The group index $n_g=c\beta_1$ (where $\beta_1$ is the first-order dispersion), and the second-order dispersion coefficient $\beta_2$ of a fibre with $\sigma_{\mathrm{av}}=1 \, \mu$m, $\Delta\sigma =0.1$, $d=0.5 \sigma$, and an output diameter $40 \sigma$ are displayed in Fig. \ref{Fig1}(a,b). The fibre effective refractive indices used in these calculations have been computed via the empirical equations \cite{Saitoh05}, including the material dispersion of silica \cite{Saleh07}.   Over the shown spectrum from 0.6 to 1 $\mu$m, the fibre dispersion is completely normal and the group index is gradually decrease with varying slopes.


Assuming   a monochromatic pump source with a wavelength 780 nm and an input power 1 W, Fig. \ref{Fig1}(c) shows the dependence of the mismatching in the propagation constants $\Delta \kappa$ of SFWM processes and the corresponding  tapering period $\Lambda_\mathrm{T}=2\pi/\Delta \kappa$   on the signal wavelength $\lambda_s$ for the average pitch.  These presented values of $\Lambda_\mathrm{T}$ are regarded as good estimates for the right  ones needed  to correct the phase-mismatching via the PTW-technique at a particular wavelength $\lambda_s$ \cite{Saleh18}.   Close to the pump, $\Lambda_\mathrm{T}$ is in the few-centimetres range, which can be  achievable using advanced post-treatment processes \cite{Mussot18}. The  PTW-technique can be used to generate photon-pairs, in principle, at any on-demand frequency. However, due to the current fabrication limitations, I will exploit this technique to demonstrate the ability to generate signal and idler photons at 750 and 812.5 nm.  All interacting photons are in the fibre fundamental mode, and their transverse profiles are approximated as Gaussian distributions \cite{Agrawal07}.

\begin{figure}
\centerline{\includegraphics[width=14.5cm]{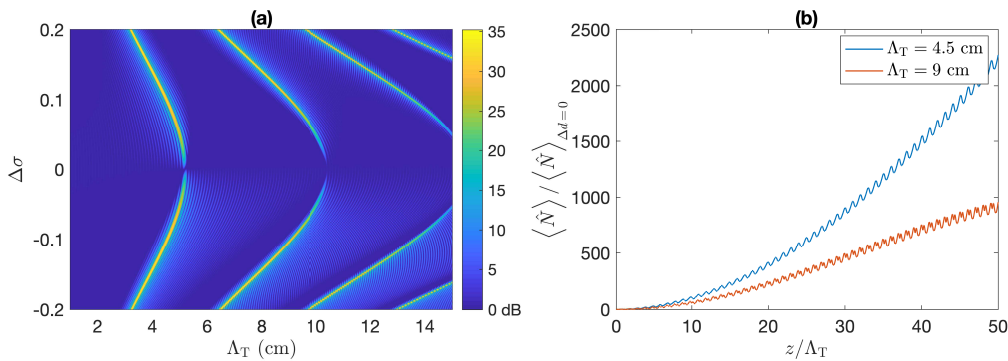}}
\caption{(a) Dependence of $\left<\hat{N}\right>$ of a SFWM process at a signal wavelength $\lambda_s$ = 750 nm on the modulation amplitude $\Delta\sigma$ and tapering period $\Lambda_\mathrm{T}$ at the end of the sinusoidally-tapered fibre, for the same number of periods $M=50$. (b) Spatial dependence of $\left<\hat{N}\right>$ with $\Delta\sigma=0.1$.
\label{Fig2}}
\end{figure}

The right combinations between the modulation amplitude and tapering period that lead to an enhanced conversion efficiency of the expected number of photons $\left<\psi\right|\hat{N}\left(L,\omega_s\right)\left|\psi\right>$ of a SFWM process at the targeted wavelengths  are portrayed in Fig. \ref{Fig2}(a) for the same number of periods $M$. The bright trajectories from left to right represent  the 1$^\mathrm{st}$, 2$^\mathrm{nd}$,3$^\mathrm{rd}$,... -order tapering periods.  The values of  $ \left<\hat{N}\right>$ are normalised to the case when   $\Delta\sigma =0$, to quantify the enhancement using the PTW-technique. The expected number of signal photons are enhanced by 35 dB  for only $M=50$. As shown Fig. \ref{Fig2}(b), the number of single photons spatially grows as an amplified oscillating function with single and double peaks within each cycle for the 1$^\mathrm{st}$- and 2$^\mathrm{nd}$-order tapering periods. Similar to periodically-poled structures, the growth rate is higher for shorter periods  \cite{Fejer92}. Also, the value of   the 2$^\mathrm{nd}$-order period slightly deviates from the exact one, which results in accumulation of small phase mismatching during propagation that subsequently affects  the growth rate.

\begin{figure}
\centerline{\includegraphics[width=14.5cm]{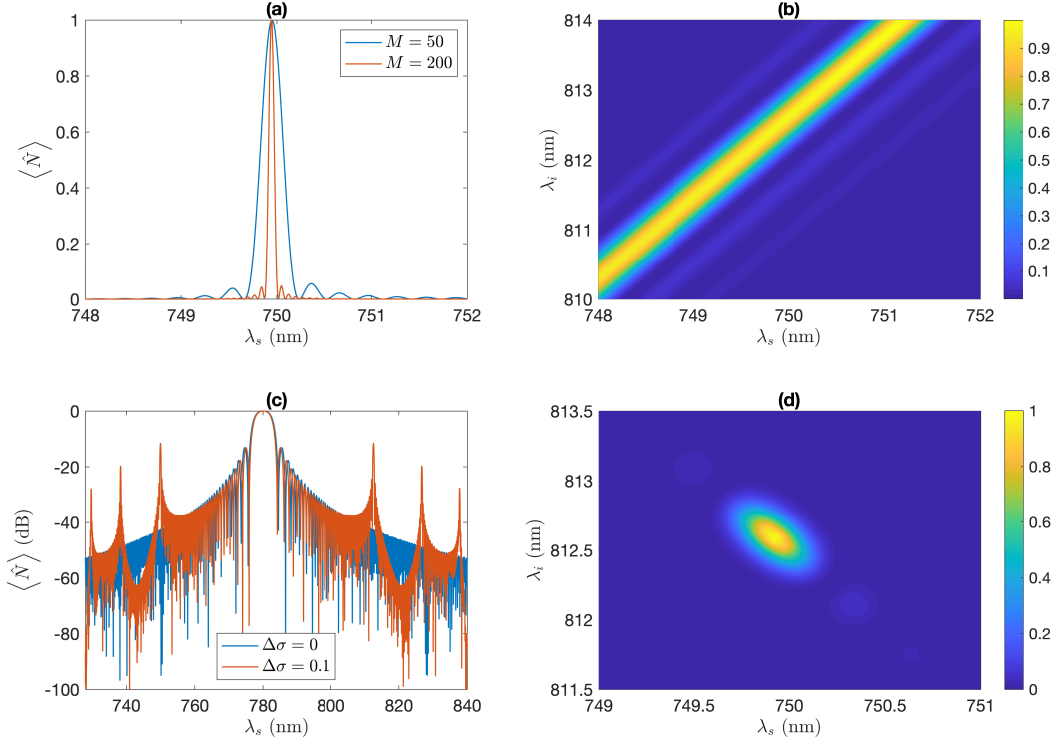}}
\caption{(a,b) 1D and 2D spectral dependence of the expected  number of photons $\left<\hat{N}\right>$ generated in a modulated fibre   with $\Delta\sigma=0.1$ around the targeted wavelengths. (c) Dependence of $\left<\hat{N}\right>$  on the signal wavelength over the entire spectrum.    (d) Dependence of $\left<\hat{N}\right>$  on the signal and idler wavelengths for a Gaussian pump with $\mathbf{E}_i$ = 1 nJ and $\tau=2.47$ ps. $M$ = 50  and $\Lambda_\mathrm{T}= $ 4.5 cm in (b--d).
\label{Fig3}}
\end{figure}

The output spectrum of the photon pairs around the targeted wavelengths at the end of the fibre is depicted in Fig. \ref{Fig3}(a) for a continuous pump wave at 780 nm. The spectrum is featured as a narrow sinc-function with very weak sidelobes that are significantly diminished for large number of periods. Panel (b) in Fig. \ref{Fig3} shows the 2D representation of the spectrum as a function of the  signal $\lambda_s$ and idler $\lambda_i$ wavelengths with  $\omega_s+\omega_i=2\omega_p$ satisfied at each point. This plot is  equivalent to the phase-matching function determined  in the interaction Hamiltonian picture via tracking the evolution of the quantum two-photon state \cite{Mosley08}.

The whole spectra in the absence and presence of the waveguide modulation  are displayed in Fig. \ref{Fig3}(c). This plot validates the developed quantum model, it resembles the output spectrum  in the normal dispersion regime of fibre due to the classical nonlinear modulation instability effect. In a uniform waveguide with a normal dispersion, the propagation of a CW is stable against small perturbations \cite{Agrawal07}. However, this scenario is completely changed in a nonuniform waveguide, where modulation instability amplifies the background noise regardless the dispersion type. The frequencies of the amplified sidebands can be approximated with  $\approx$ 3 nm difference from the simulations using the relation $\beta_2\Omega^2+2 \gamma_{s, i}P=2\pi l/\Lambda_\mathrm{T}$, where $\Omega=\omega_s-\omega_p$ is the frequency shift from the pump source, $\beta_2$, and $\gamma_{s, i}$ are the average second-order dispersion and nonlinear coefficient,  $P$ is the input power, and $l$ is an integer \cite{Mussot18}. The slight deviation is due to the negligence of the effect of the modulation amplitude  in this  equation.

Considering a Gaussian-pulse pump source with an input energy 1 nJ and a full-width-half-maximum 4 ps, the corresponding expected number of photons is portrayed in Fig. \ref{Fig3}(d). This plot is  equivalent to the joint-spectral intensity distribution, which is usually obtained in the Hamiltonian picture and regarded as the magnitude square of the product of the phase-matching function and the pump-spectral envelope \cite{Dosseva16}. Therefore, the joint-spectral amplitude function, used in calculating the spectral-purity  or discorrelation between the output single photons,  can be  constructed from the elements $\mathbf{T}_{s,i}\left(1,2\right)$ in the introduced model. The  group-velocity-matching condition, required for high spectral-purity  \cite{Garay-Palmett07}, is satisfied  in the simulated case via setting the  group-velocity of the  pump in between those of the signal and idler,  as shown in  Fig. \ref{Fig1}(a). Using the Schmidt decomposition  of the  matrix constructed from the elements $\mathbf{T}_{s,i}\left(1,2\right)$ results in a spectral purity 0.74  \cite{Mosley08}.  For a shorter Gaussian pump source with width 1 ps the spectral purity increases to 0.93, which shows the ability of the PTW-technique in producing high-pure single photons at any sought frequencies.   

\subsection{Width-modulated silicon-nitride waveguides.}
\begin{figure}
\centerline{\includegraphics[width=14.5cm]{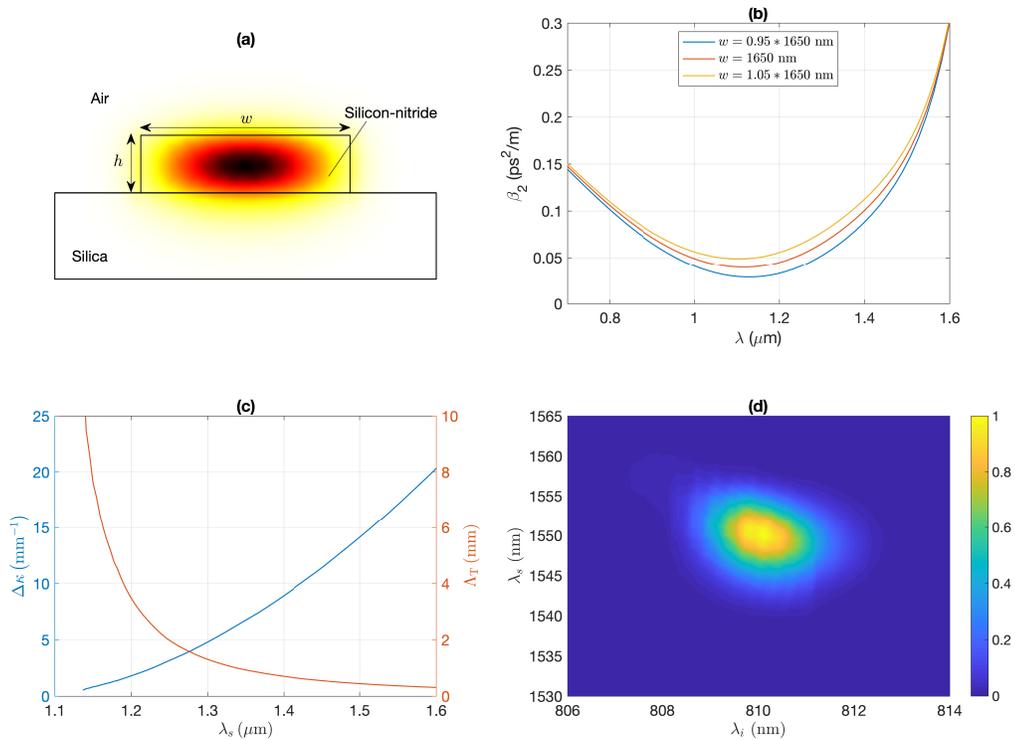}}
\caption{ A planar silicon-nitride PTW  with $w_{\mathrm{av}}=1650 $ nm,  $\Delta w = 0.05$, $h=450$ nm,   $\Lambda_\mathrm{T}=370\,\mu$m,  $M =50$,  and  $n_2=1.5\times 10^{-18}$ m$^2$/W: (a) Fundamental TE mode at $\lambda = 1064\, $nm and  $w=w_{\mathrm{av}}$. (b) Spectral-dependence of $\beta_2$ of the waveguide at different widths. (c) Dependence of  $\Delta\kappa$ and the corresponding  $\Lambda_\mathrm{T}$ on the signal wavelength $\lambda_s$  at the average width, and a pump wavelength at 1064 nm with an input power 1 W. (d) Dependence of $\left<\hat{N}\right>$  on the signal and idler wavelengths for a Gaussian-pulse centred at 1064 nm with $\mathbf{E}_i$ = 1 nJ and $\tau=0.29$ ps.
\label{Fig4}}
\end{figure}

A current key challenge in developing integrated single photons sources is isolating the pump photons from the converted single photons that are only within few nanometers spectral range \cite{Pavesi16}.  The PTW-technique would provide a direct solution for this problem by enabling photon-pairs generation at on-demand frequencies spectrally far from the pump. For this purpose, planar silicon-nitride waveguides, characterised by CMOS-compatibility, negligible two-photon absorption and Raman nonlinearity \cite{Guo18},  have been also examined in this work. A sketch of the cross-sectional view of the planar waveguide used in the simulations is shown in  Fig. \ref{Fig4}(a). In this case, the core thickness $h$  is fixed, while its width varies sinusoidally as  $w(z)=w_{\mathrm{av}}\left[1-\Delta w\,\cos\left( 2\pi z/\Lambda_\mathrm{T}\right)\right]$ with $w_{\mathrm{av}}$ the average width and $\Delta w$ the modulation. The  fundamental TE mode of a wave at 1.064 $\mu$m in a  waveguide with a core width $w=1650 $ nm and a thickness $h=450$ nm is shown in Fig. \ref{Fig4}(a). Simulations are performed using  `COMSOL', a commercial finite-element method,  including the material dispersion of silica and silicon nitride. The dispersion of the waveguide is normal over the entire spectrum, as depicted in  Fig. \ref{Fig4}(b). The guiding loss is less than 0.1 dB/m for $\lambda < 1.4 \,\mu$m, then it increases rapidly to 3.8 dB/m at 1.6 $\mu$m. 

Assuming a pump source with a central wavelength 1.064 $\mu$m and an input pump power 1 W, Fig. \ref{Fig4}(c) shows the spectral dependence of the propagation-constant mismatching $\Delta \kappa$  of SFWM processes and the corresponding  tapering periods for the average waveguide width. $\Lambda_\mathrm{T} $ in the range of few-hundreds micrometers  is required to generate photon-pairs with one photon in the telecommunication regime using this pump source. The expected number of photons in a PTW  with 50 periods under a Gaussian-pulse excitation is depicted in  Fig. \ref{Fig4}(d). The emission of photon-pairs at 810 and 1550 nm has been enhanced by 25 dB via the proper combination of the  tapering period and modulation amplitude. The presented data has been smoothed to amend small irregularities, which might be due to the fact that an extremely-fine mesh grid is required to precisely resolve the modes  of this planar configuration especially  at longer wavelengths. The spectral purity of the output photons is 0.82 (0.9 with smoothing) using Schmidt decomposition, which demonstrates the potential of the proposed structure as a pure highly-efficient single-photon source for integrated quantum applications.

\section{Conclusions}\label{Sec4}
In conclusion, I have developed a rigorous quantum model to study SFWM parametric processes in tapered waveguides for continuous and pulsed pump sources. The model calculates  the expected number photons of a certain spectral mode in the Heisenberg picture.  The transverse profiles of the interacting modes,  material and waveguide dispersions, self and cross-phase modulations have been accurately treated in this analysis. Both dispersion-oscillating fibres and width-modulated  silicon-nitride waveguides, examples of PTWs,  have been examined for enhancing photon-pairs generation at on-demand frequencies and with high spectral purities. The model results have  been verified via retrieving the dynamics of the classical nonlinear modulation instability process in dispersion-oscillating fibres.  In comparison to uniform waveguides, a 35 (25) dB enhancement in the expected number of photons can be achieved via a proper choice of the tapering period and modulation amplitude in fibres (planar waveguides) with only 50 periods. The output photons are characterised by having very narrow bandwidths, and high spectral purities. The presented  simulations theoretically demonstrate the concept of the PTW-technique as a new efficient quasi-phase-matching scheme for photon-pairs generation in third-order nonlinear materials, which would have a great leverage in the field of quantum optics. Finally, I anticipate that exploring non-periodic tapering patterns could be a fruitful future-research direction for optimising the spectral properties of the output photons. 

\section*{Acknowledgment }
The author thanks F. Graffitti and  Dr. A. Fedrizzi at Heriot-Watt University for useful discussions.

\section*{Funding}
 Royal Society of Edinburgh (RSE) (501100000332).


\begin{thebibliography}{10}
\newcommand{\enquote}[1]{``#1''}

\bibitem{Spring17}
J.~B. Spring, P.~L. Mennea, B.~J. Metcalf, P.~C. Humphreys, J.~C. Gates, H.~L.
  Rogers, C.~S\"{o}ller, B.~J. Smith, W.~S. Kolthammer, P.~G.~R. Smith, and
  I.~A. Walmsley, \enquote{Chip-based array of near-identical, pure, heralded
  single-photon sources,} Optica \textbf{4}, 90--96 (2017).

\bibitem{Armstrong62}
J.~A. Armstrong, N.~Bloembergen, J.~Ducuing, and P.~S. Pershan,
  \enquote{Interactions between light waves in a nonlinear dielectric,} Phys.
  Rev. \textbf{127}, 1918--1939 (1962).

\bibitem{Fejer92}
M.~M. Fejer, G.~A. Magel, D.~H. Jundt, and R.~L. Byer,
  \enquote{Quasi-phase-matched second harmonic generation: tuning and
  tolerances,} IEEE Journal of Quantum Electronics \textbf{28}, 2631--2654
  (1992).

\bibitem{Eckstein11}
A.~Eckstein, A.~Christ, P.~J. Mosley, and C.~Silberhorn, \enquote{Highly
  efficient single-pass source of pulsed single-mode twin beams of light,}
  Phys. Rev. Lett. \textbf{106}, 013603 (2011).

\bibitem{Spring13}
J.~B. Spring, P.~S. Salter, B.~J. Metcalf, P.~C. Humphreys, M.~Moore,
  N.~Thomas-Peter, M.~Barbieri, X.-M. Jin, N.~K. Langford, W.~S. Kolthammer,
  M.~J. Booth, and I.~A. Walmsley, \enquote{On-chip low loss heralded source of
  pure single photons,} Opt. Express \textbf{21}, 13522--13532 (2013).

\bibitem{Graffitti18}
F.~Graffitti, P.~Barrow, M.~Proietti, D.~Kundys, and A.~Fedrizzi,
  \enquote{Independent high-purity photons created in domain-engineered
  crystals,} Optica \textbf{5}, 514--517 (2018).

\bibitem{McMillan09}
A.~McMillan, J.~Fulconis, M.~Halder, C.~Xiong, J.~Rarity, and W.~Wadsworth,
  \enquote{Narrowband high-fidelity all-fibre source of heralded single photons
  at 1570 nm,} Opt. Express \textbf{17}, 6156--6165 (2009).

\bibitem{Silverstone13}
J.~W. Silverstone, D.~Bonneau, K.~Ohira, N.~Suzuki, H.~Yoshida, N.~Iizuka,
  M.~Ezaki, C.~M. Natarajan, M.~G. Tanner, R.~H. Hadfield, V.~Zwiller, G.~D.
  Marshall, J.~G. Rarity, J.~L. O'Brien, and M.~G. Thompson, \enquote{On-chip
  quantum interference between silicon photon-pair sources,} Nature Photonics
  \textbf{8}, 104 EP -- (2013).

\bibitem{Agrawal07}
G.~P. Agrawal, \emph{Nonlinear Fiber Optics}, San Diego, California (Academic
  Press, 2007), 4th ed.

\bibitem{Rarity05}
J.~G. Rarity, J.~Fulconis, J.~Duligall, W.~J. Wadsworth, and P.~S.~J. Russell,
  \enquote{Photonic crystal fiber source of correlated photon pairs,} Opt.
  Express \textbf{13}, 534--544 (2005).

\bibitem{Garay-Palmett07}
K.~Garay-Palmett, H.~J. McGuinness, O.~Cohen, J.~S. Lundeen, R.~Rangel-Rojo,
  A.~B. U'Ren, M.~G. Raymer, C.~J. McKinstrie, S.~Radic, and I.~A. Walmsley,
  \enquote{Photon pair-state preparation with tailored spectral properties by
  spontaneous four-wave mixing in photonic-crystal fiber,} Opt. Express
  \textbf{15}, 14870--14886 (2007).

\bibitem{Vernon17}
Z.~Vernon, M.~Menotti, C.~C. Tison, J.~A. Steidle, M.~L. Fanto, P.~M. Thomas,
  S.~F. Preble, A.~M. Smith, P.~M. Alsing, M.~Liscidini, and J.~E. Sipe,
  \enquote{Truly unentangled photon pairs without spectral filtering,} Opt.
  Lett. \textbf{42}, 3638--3641 (2017).

\bibitem{Dong04}
P.~Dong and A.~G. Kirk, \enquote{Nonlinear frequency conversion in waveguide
  directional couplers,} Phys. Rev. Lett. \textbf{93}, 133901 (2004).

\bibitem{Jones18}
R.~J.~A. Francis-Jones, T.~A. Wright, A.~V. Gorbach, and P.~J. Mosley,
  \enquote{Engineered photon-pair generation by four-wave mixing in asymmetric
  coupled waveguides,} arXiv:1809.10494  (2018).

\bibitem{Driscoll12}
J.~B. Driscoll, N.~Ophir, R.~R. Grote, J.~I. Dadap, N.~C. Panoiu, K.~Bergman,
  and R.~M. Osgood, \enquote{Width-modulation of si photonic wires for
  quasi-phase-matching of four-wave-mixing: experimental and theoretical
  demonstration,} Opt. Express \textbf{20}, 9227--9242 (2012).

\bibitem{Mussot18}
A.~Mussot, M.~Conforti, S.~Trillo, F.~Copie, and A.~Kudlinski,
  \enquote{Modulation instability in dispersion oscillating fibers,} Adv. Opt.
  Photon. \textbf{10}, 1--42 (2018).

\bibitem{Hickstein18}
D.~D. Hickstein, G.~C. Kerber, D.~R. Carlson, L.~Chang, D.~Westly,
  K.~Srinivasan, A.~Kowligy, J.~E. Bowers, S.~A. Diddams, and S.~B. Papp,
  \enquote{Quasi-phase-matched supercontinuum generation in photonic
  waveguides,} Phys. Rev. Lett. \textbf{120}, 053903 (2018).

\bibitem{Saleh18}
M.~F. Saleh, \enquote{Quasi-phase-matched
  ${\ensuremath{\chi}}^{(3)}$-parametric interactions in sinusoidally tapered
  waveguides,} Phys. Rev. A \textbf{97}, 013850 (2018).

\bibitem{Huttner90}
B.~Huttner, S.~Serulnik, and Y.~Ben-Aryeh, \enquote{Quantum analysis of light
  propagation in a parametric amplifier,} Phys. Rev. A \textbf{42}, 5594--5600
  (1990).

\bibitem{Guo18}
H.~Guo, C.~Herkommer, A.~Billat, D.~Grassani, C.~Zhang, M.~H.~P. Pfeiffer,
  W.~Weng, C.-S. Br{\`e}s, and T.~J. Kippenberg, \enquote{Mid-infrared
  frequency comb via coherent dispersive wave generation in silicon nitride
  nanophotonic waveguides,} Nature Photonics \textbf{12}, 330--335 (2018).

\bibitem{Saleh11a}
M.~F. Saleh, W.~Chang, P.~H\"olzer, A.~Nazarkin, J.~C. Travers, N.~Y. Joly,
  P.~S.~J. Russell, and F.~Biancalana, \enquote{Theory of
  photoionization-induced blueshift of ultrashort solitons in gas-filled
  hollow-core photonic crystal fibers,} Phys. Rev. Lett. \textbf{107}, 203902
  (2011).

\bibitem{Love91}
J.~D. Love, W.~M. Henry, W.~J. Stewart, R.~J. Black, S.~Lacroix, and
  F.~Gonthier, \enquote{Tapered single-mode fibres and devices. i. adiabaticity
  criteria,} IEE Proceedings J - Optoelectronics \textbf{138}, 343--354 (1991).

\bibitem{Kowligy18}
A.~S. Kowligy, D.~D. Hickstein, A.~Lind, D.~R. Carlson, H.~Timmers, N.~Nader,
  D.~L. Maser, D.~Westly, K.~Srinivasan, S.~B. Papp, and S.~A. Diddams,
  \enquote{Tunable mid-infrared generation via wide-band four-wave mixing in
  silicon nitride waveguides,} Opt. Lett. \textbf{43}, 4220--4223 (2018).

\bibitem{Saitoh05}
K.~Saitoh and M.~Koshiba, \enquote{Empirical relations for simple design of
  photonic crystal fibers,} Opt. Express \textbf{13}, 267--274 (2005).

\bibitem{Saleh07}
B.~E.~A. Saleh and M.~C. Teich, \emph{Fundamentals of Photonics} (Wiley, 2007),
  2nd ed.

\bibitem{Mosley08}
P.~J. Mosley, J.~S. Lundeen, B.~J. Smith, and I.~A. Walmsley,
  \enquote{Conditional preparation of single photons using parametric
  downconversion: a recipe for purity,} New Journal of Physics \textbf{10},
  093011 (2008).

\bibitem{Dosseva16}
A.~Dosseva, L.~Cincio, and A.~M. Bra\ifmmode~\acute{n}\else \'{n}\fi{}czyk,
  \enquote{Shaping the joint spectrum of down-converted photons through
  optimized custom poling,} Phys. Rev. A \textbf{93}, 013801 (2016).

\bibitem{Pavesi16}
L.~Pavesi and D.~J. Lockwood, \emph{Silicon Photonics III} (Springer-Verlag
  Berlin Heidelberg, 2016).

\end{thebibliography}
\end{document}